\documentclass[sigconf]{acmart}
\AtBeginDocument{%
  \providecommand\BibTeX{{%
    \normalfont B\kern-0.5em{\scshape i\kern-0.25em b}\kern-0.8em\TeX}}}

\copyrightyear{2022}
\acmYear{2022}
\setcopyright{acmcopyright}\acmConference[WWW '22 Companion]{Companion Proceedings of the Web Conference 2022}{April 25--29, 2022}{Virtual Event, Lyon, France}
\acmBooktitle{Companion Proceedings of the Web Conference 2022 (WWW '22 Companion), April 25--29, 2022, Virtual Event, Lyon, France}
\acmPrice{15.00}
\acmDOI{10.1145/3487553.3524215}
\acmISBN{978-1-4503-9130-6/22/04}

\begin{document}

\title[Beyond NDCG]{Beyond NDCG: behavioral testing of recommender\\systems with RecList}

%% CFP: https://www2022.thewebconf.org/cfp/industry/

%% original doc: https://docs.google.com/document/d/12lFuwuOjVMdCdCbqLuhr0h2h9KqLp2EmuPcGHotOSnI/edit

\author{Patrick John Chia}
\authornote{Patrick, Jacopo and Federico originally conceived and designed \texttt{RecList} together, and they contributed equally to the paper. Chloe and Brian added important capabilities to the package, and greatly helped in improving the paper as well.}
\affiliation{%
  \institution{Coveo}
  \country{Canada}
}
\email{pchia@coveo.com}

\author{Jacopo Tagliabue}
\affiliation{%
  \institution{Coveo Labs}
  \country{United States}}
\email{jtagliabue@coveo.com}

\author{Federico Bianchi}
\affiliation{%
  \institution{Bocconi University}
  \country{Italy}
}
\email{f.bianchi@unibocconi.it}

\author{Chloe He}
\affiliation{%
  \institution{Stanford University}
  \country{United States}
}
\email{chloehe@stanford.edu}

\author{Brian Ko}
\affiliation{%
  \institution{KOSA AI}
  \country{United States}
}
\email{sangwoo@kosa.ai}

\renewcommand{\shortauthors}{Chia et al.}

\begin{abstract}
  As with most Machine Learning systems, recommender systems are typically evaluated through performance metrics computed over held-out data points. However, real-world behavior is undoubtedly nuanced: \textit{ad hoc} error analysis and tests must be employed to ensure the desired quality in actual deployments. We introduce \texttt{RecList}, a testing methodology providing a general plug-and-play framework to scale up behavioral testing. We demonstrate its capabilities by analyzing known algorithms and black-box APIs, and we release it as an open source, extensible package for the community.
\end{abstract}


\begin{CCSXML}
<ccs2012>
<concept>
<concept_id>10011007.10011074.10011099.10011105.10011109</concept_id>
<concept_desc>Software and its engineering~Acceptance testing</concept_desc>
<concept_significance>300</concept_significance>
</concept>
<concept>
<concept_id>10002951.10003317.10003347.10003350</concept_id>
<concept_desc>Information systems~Recommender systems</concept_desc>
<concept_significance>500</concept_significance>
</concept>
</ccs2012>
\end{CCSXML}

\ccsdesc[300]{Software and its engineering~Acceptance testing}
\ccsdesc[500]{Information systems~Recommender systems}

\keywords{recommender systems, behavioral testing, open source}

\maketitle

\section{Introduction}

\begin{quote}
``A QA engineer walks into a bar. Orders a beer. Orders 0 beers. Orders 99999999999 beers. Orders a lizard. Orders -1 beers. Orders a ueicbksjdhd. First real customer walks in and asks where the bathroom is. The bar bursts into flames, killing everyone'' -- B. Keller (random tweet).
\end{quote}

In recent years, recommender systems (hence \textbf{RSs}) have played an indispensable role in providing personalized digital experiences to users, by fighting information overload and helping with navigating inventories often made of millions of items~\cite{Chia2021AreYS,10.1145/3219819.3219891,nassif2018diversifying,DBLP:journals/corr/abs-2005-10110,wang2019deconfounded}. RSs' ability to generalize, both in industry and academia, is often evaluated through some accuracy score over a held-out dataset: however, performance given by a single number often fails to give developers and stakeholders a rounded view of the expected performances of the system ``in the wild''. For example, as industry seems to recognize more than academia, not all inputs are created equal, and not all mistakes are uniformly costly; while these considerations are crucial to real-world success, reporting NDCG alone fails to capture these nuances. This is particularly important in the world of RSs, given both the growing market for RSs\footnote{E-commerce alone -- arguably the biggest market for recommendations -- is estimated to turn into a $>4$ trillion industry by the end of 2021~\cite{ecomworld}.} and the role of RSs in shaping (often, narrowing~\cite{concentrationbias}) user preferences with potential harmful consequences~\cite{netflixarticle}.

Following the lead of \cite{Ribeiro2020BeyondAB} in Natural Language Processing, we propose a behavioral-based framework to test RSs across a variety of industries, focusing on the peculiarities of horizontal use cases (e.g. substitute vs complementary items) more than vertical domains. We summarize our main contributions as follows:

\begin{itemize}
    \item we argue for the importance of a well-rounded and more nuanced evaluation of RSs and discuss the importance of scaling up testing effort through automation;
    \item we release an open-source package to the community -- \texttt{RecList}. \texttt{RecList} comes with ready-made behavioral tests and connectors for important public datasets (\textit{Coveo Data Challenge} \cite{CoveoSIGIR2021}, \textit{MovieLens} \cite{10.1145/2827872}, \textit{Spotify} \cite{10.1145/3344257}) and an extensible interface for custom use cases;
    \item we demonstrate our methodology by analyzing standard models and SaaS offerings over a cart recommendation task. 
\end{itemize}

While we developed \texttt{RecList} out of the very practical necessities involved in scaling RSs to hundreds of organizations across many industries\footnote{\textit{Coveo} is a \textit{multi-tenant} provider of A.I. services, with a network of hundreds of deployments for customer service, e-commerce and enterprise search use cases.}, as researchers,  we also believe this methodology to be widely applicable in error analysis and thorough evaluation of new models: as much as we like to read about a new SOTA score on \textit{MovieLens}, we would also like to understand what that score tells us about the capabilities and shortcomings of the model.

\section{An Industry Perspective}
\label{sec:industry}

While quantitative metrics over standardized datasets are indispensable to provide an objective pulse on where the field is going, we often find that NDCG tells only one part of the performance story. As a very concrete example, while model performance depends mostly on what happens with frequent items, the final user experience may be ruined by poor outcomes in the long-tail~\cite{mckinsey}. Metrics such as coverage, serendipity, and bias \cite{Kotkov2016ChallengesOS,jannach2017recurrent,ludewig2018evaluation} have been proposed to capture other aspects of the behaviors of RSs, but they still fall short of what is needed to debug RSs in production, and often do not provide any guarantee that a model will be reliable when released.

\begin{figure}[h]
  \centering
  \includegraphics[width=\linewidth]{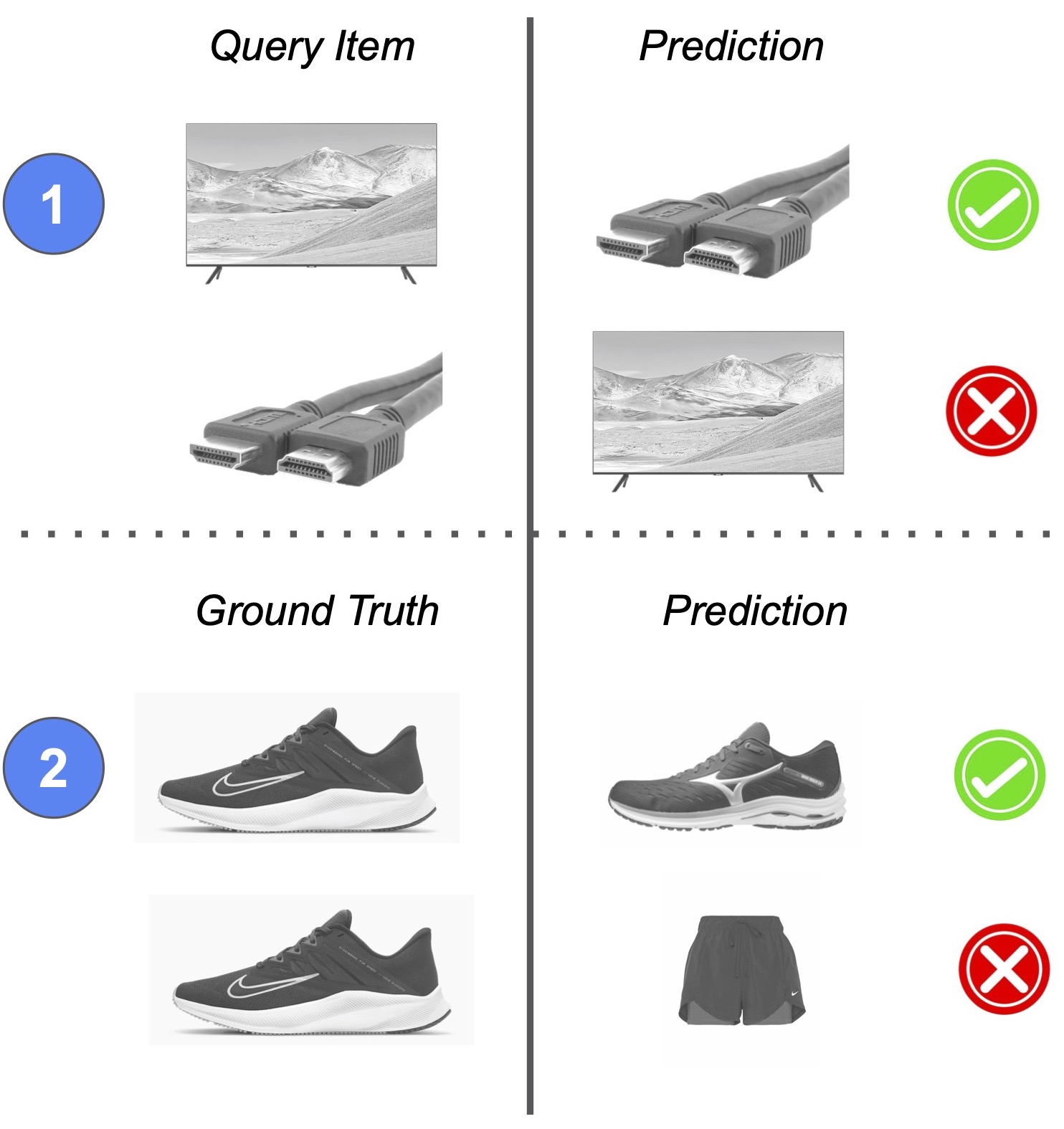}
  \caption{Examples of behavioral principles for RSs: in (1) we observe the asymmetry desired when recommending complementary items, while (2) exemplifies that model mistakes (i.e. missing the ground truth item) may degrade the shopping experience in different ways.}
  \label{fig:tests}
\end{figure}

When developing \texttt{RecList}, we started from popular use cases that represent the most widely adopted strategies for recommendation systems:

\begin{enumerate}
    \item \textbf{similar items}: when shown running shoes, users may want to browse for another pair of running shoes -- in other words, they are looking for substitutable products \cite{Chia2021AreYS}; similarly, in entertainment \cite{nassif2018diversifying,lamkhede2021recommendations} RSs may suggest content similar to a previous viewing;
    \item \textbf{complementary items}: when a TV has been added to the cart, shoppers may want to buy a complementary product (e.g. a cable). This type of recommendation is typical of e-commerce scenarios and exhibits a characteristic asymmetry (Figure~\ref{fig:tests}); 
    \item\textbf{session-based recommendations}: real-time behavior has been recently exploited to provide session-based personalization ~\cite{tagliabue-etal-2021-bert,Wang2019ASO,10.1145/3292500.3330839,Hidasi2016SessionbasedRW}, which captures both preferences from recent sessions and real-time intent; a typical session-based RS ingests the latest item interactions for a user and predicts the next interaction(s). 
\end{enumerate}

From these use cases, we identified three main areas of behavioral intervention:

\begin{enumerate}
    \item \textbf{enforce per-task invariants}: irrespective of the target deployment, complementary and similar items satisfy formal relations which are different in nature. In particular, similar items need to be interchangeable, while complementary items may have a natural ordering (See Fig.~\ref{fig:tests}). We operationalize these insights by joining predictions with item metadata: for example, we can use price information to check for asymmetry constraints;
    \item \textbf{being less wrong}: if the ground truth item for a movie recommendation is ``When Harry met Sally'', hit-or-miss metrics won't be able to distinguish between model A that predicts ``Terminator'' and model B that suggests ``You've got mail''\footnote{In case the reader is too young to know better, suggesting ``Terminator'' in this context is way worse than suggesting ``You've got mail''.}. In other words, A and B are not wrong in the same way: one is a terrible suggestion and one is reasonable mistake. RSs are a major factor in boosting user experience (which translates to revenues, loyalty, etc.): in a recent survey, 38\% of shoppers said they would stop shopping if shown non-relevant recommendations \cite{emarketer};
    \item\textbf{data slices}: in real-world RSs, not all inputs are created equal. In particular, we may tolerate a small decrease in overall accuracy if a subset of users we care about is happier. For a practical example, consider a multi-brand retailer promoting the latest Nike shoes with a marketing campaign: other things being equal, this retailer would want to make sure the experiences of users landing on Nike product pages are particularly well curated. Aside from horizontal cases (e.g. \textit{cold-start} items), the most interesting slices are often context-dependent, which is an important guiding principle for our library.
\end{enumerate}

Building \texttt{RecList} requires us to solve two problems: operationalize behavioral principles in code whenever possible, \textit{and} provide an extensible interface when domain knowledge and custom logic are required (Section~\ref{sec:software}).

\section{Related work}
This work sits at the intersection of several themes in the research and industrial communities. We were initially inspired by behavioral testing for NLP pioneered by \cite{Ribeiro2020BeyondAB}: from this seminal work we took two lessons: \textit{first}, that black-box testing \cite{536464} is a source of great insights when added to standard metrics; \textit{second}, that this methodology goes hand-in-hand with software tools, as creating, maintaining, and analyzing behavioral tests by manual curation is a time-consuming process. On the other hand, \texttt{RecList} needs to consider the peculiarities of RSs, as compared to NLP: in particular, the concept of generic models does not apply, as RSs are deployed in different shops and domains: the same pair of running shoes can be popular in \textit{Shop X} and not \textit{Shop Y}, and categorized as \textit{sneakers} in one case, \textit{running shoes} in the other.

From the A/B testing literature \cite{10.1145/2339530.2339653}, we take the important lesson that not all test cases are created equal: in particular, just as a careful A/B test cares both about the aggregate effect of treatment and the individual effects on specific data slices, a careful set of RS testing should worry about the overall accuracy as well as the accuracy in specific subgroup-of-interests: in ML systems, as in life, gains and losses are not always immediately interchangeable.%\footnote{A common situation with the conditional average treatment effect is realizing that treatment has better overall accuracy but it is making the experience of an important group significantly worse.}.

The RS literature exploited already insights contained in \texttt{RecList}, typically as part of error analysis \cite{10.1145/3460231.3474620}, or as performance boost for specific datasets \cite{Moreira2021TransformersWM}. For example, ``being less wrong'' is discussed in \cite{10.1145/3383313.3411477}, while cold start performance is often highlighted for methods exploiting content-based features \cite{10.1145/2959100.2959160}. Our work builds on top of this scattered evidence, and aims to be the one-stop shop for behavioral analysis of RSs: \texttt{RecList} provides practitioners with both a common lexicon and working code for scalable, in-depth error analysis.

Finally, as far as standard metrics go, the literature is pretty consistent: a quick scan through recent editions of RecSys and SIGIR highlights the use of \textit{MRR}, \textit{ACCURACY}, \textit{HITS}, \textit{NDCG} as the main metrics~\cite{wang2019neural,Rashed2020MultiRecAM,10.1145/3383313.3412235,10.1145/3383313.3412263,10.1145/3404835.3462832}. To ease the comparison with research papers on standard KPIs, we made sure that these metrics are computed by \texttt{RecList} as well, together with behavioral results.

\section{RecList (a.k.a. \texttt{CheckList} for recs)}
\label{sec:software}

\begin{figure*}
  \centering
  \includegraphics[width=\textwidth]{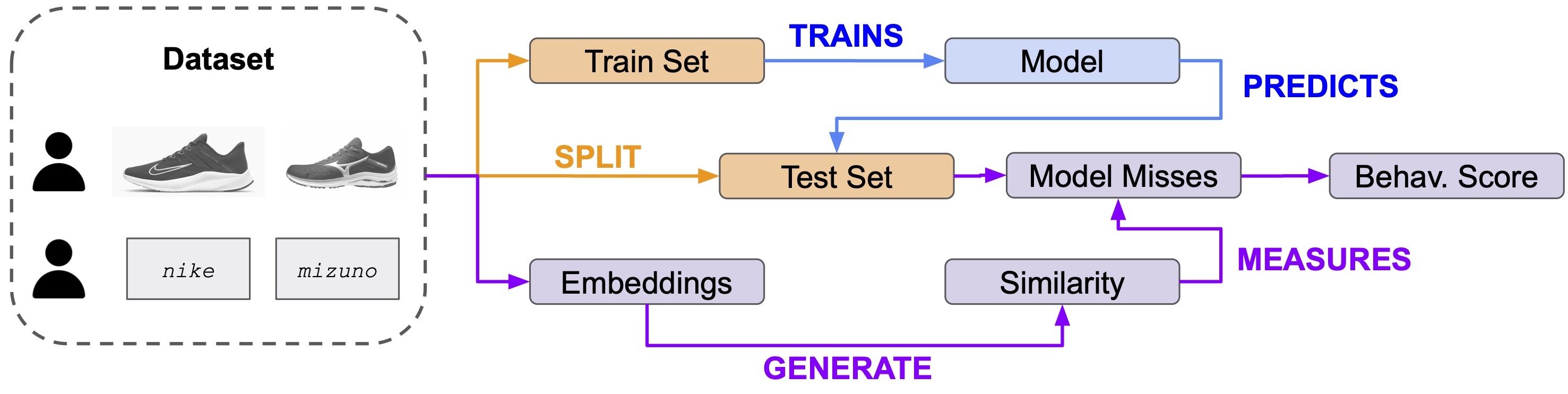}
  \caption{Sample workflow for behavioral tests. Starting with shopping data (left), the dataset split (orange) and model training (blue) mimic the usual training loop. \texttt{RecList} creates a latent space to measure the relationships between inputs, ground truths and predictions, such as how far misses are from ground truths (violet) (see Fig.~\ref{fig:query_dist} for a real-world example). Since a session can be viewed as a sequence of items or features (brands), \texttt{RecList} can re-use skip-gram to create embeddings for different tests.}
  \label{fig:perturbation}
  \vspace{-4mm}
\end{figure*}

\texttt{RecList} is behavioral testing applied to RSs, and available as a plug-and-play open-source package that can be easily extended to proprietary datasets and models. Following~\cite{Ribeiro2020BeyondAB}, we decouple testing from implementation: our framework treats RSs as a black box (through an extensible programming interface), allowing us to test RSs for which no source code is available (e.g. SaaS models). To strengthen our exposition of the methodology, we offer here a high-level view of the logical architecture and capabilities of \texttt{RecList} as a package. However, please note the code is actively evolving as a community project: the reader is encouraged to check out our repository\footnote{\url{https://github.com/jacopotagliabue/reclist}} for up-to-date documentation, in-depth explanation of available tests and practical examples over popular datasets and baseline models.

\subsection{Abstractions}
\texttt{RecList} is a Python package built over these main abstractions: 

\begin{itemize}
    \item \textit{RecTask}: the recommendation use case (Section~\ref{sec:industry}).
    \item \textit{RecModel}: the model we are testing -- as long as a simple prediction-based interface can be implemented, any model can be represented in \texttt{RecList}. For example, a SaaS model would make an API call to a service and let \texttt{RecList} handle the analysis. 
    \item \textit{RecDataset}: the dataset we are using -- the class provides standard access to train/test splits \textit{and} item metadata. \texttt{RecList} comes with ready-made connectors for popular datasets. 
    \item \textit{RecList}\footnote{Note that we use \textit{RecList} to indicate the class or its instances, and \texttt{RecList} to indicate the package as a whole.}: the actual set of tests we are running, given a \textit{RecTask}, \textit{RecModel} and \textit{RecDataset}. A \textit{RecList} is made of \textit{RecTests}.
\end{itemize}

When running a \textit{RecList}, the package automatically versions the relevant metadata: results are exported in a machine-friendly format, and can be easily ingested in existing ML tools \cite{CometML,Metaflow} to visually compare the performance of different models.

\subsection{Capabilities}
\label{sec:tests}
While we refer readers to our repository for an up-to-date list of available \textit{RecLists}, \textit{RecModels} and \textit{RecDatasets}, we wish to highlight some key capabilities:

\begin{itemize}
    \item \textbf{leveraging representation learning}: word embeddings for behavioral testing in NLP are replaced by representational learning \textit{per dataset}. By unifying access to items and metadata (e.g. \textit{brands} for products, \textit{labels} for music), \texttt{RecList} provides a scalable, unsupervised flow to obtain latent representation of target entities, and uses them to generate new test pairs, or supply similarity judgment when needed (Figure~\ref{fig:perturbation}). \texttt{RecList} ships with \textit{prod2vec} over session-like data \cite{Bianchi2020FantasticEA,Mikolov2013EfficientEO}, but the same idea would work with other representational techniques (e.g. zero-shot representations \cite{Radford2021LearningTV}, BERT-based embeddings \cite{tagliabue-etal-2021-bert});
    \item \textbf{merging metadata and predictions}: \texttt{RecList}'s tests provide a functional interface that can be applied to any dataset by supplying the corresponding entities. For example, asymmetry tests can be applied to any feature exhibiting the desired behavior (e.g. \textit{price} for complementary items); in the same vein, data slices can be specified with arbitrary partitioning functions, allowing seamless reporting on important subsets;
    \item \textbf{injecting domain knowledge when needed}: \texttt{RecList} allows to easily swap default similarity metrics with custom ones (or, of course, write entirely new tests): for example, if a practitioner is working in a domain with a very accurate taxonomy, he could define a new distance between predictions and labels, supplementing out-of-the-box unsupervised similarity metrics.
\end{itemize}

\begin{table}[h]

  \caption{Results for a complementary \textit{RecList}.}
  \label{tab:results}
  \begin{tabular}{l|cccc}
    \toprule
    \textbf{Test}&\textbf{P2V}&\textbf{GOO}&\textbf{S1}                    \\
    \midrule
    HR@10                  & 0.197              & \textbf{0.199}   & 0.094    \\
    MRR@10                 & 0.091             & \textbf{0.102}   & 0.069    \\
    Coverage@10            & 1.01e-2            & \textbf{1.99-e2} & 3.00e-3  \\
    Popularity Bias@10     & \textbf{9.91e-5}   & 1.41e-4          & 1.20e-4   \\
    \midrule
    Cos Distance (Brand)   & \textbf{0.411}  & 0.483        & 0.540 \\
    Cos Distance (Misses)  & 0.564 & \textbf{0.537}  & 0.577      \\
    Path Length (Category) & 1.13   & 1.59   & \textbf{1.91} \\
  \bottomrule
\end{tabular}
\end{table}

\section{A Worked-out example: cart recs}
\begin{figure}[h]
  \centering
  \includegraphics[scale=0.51]{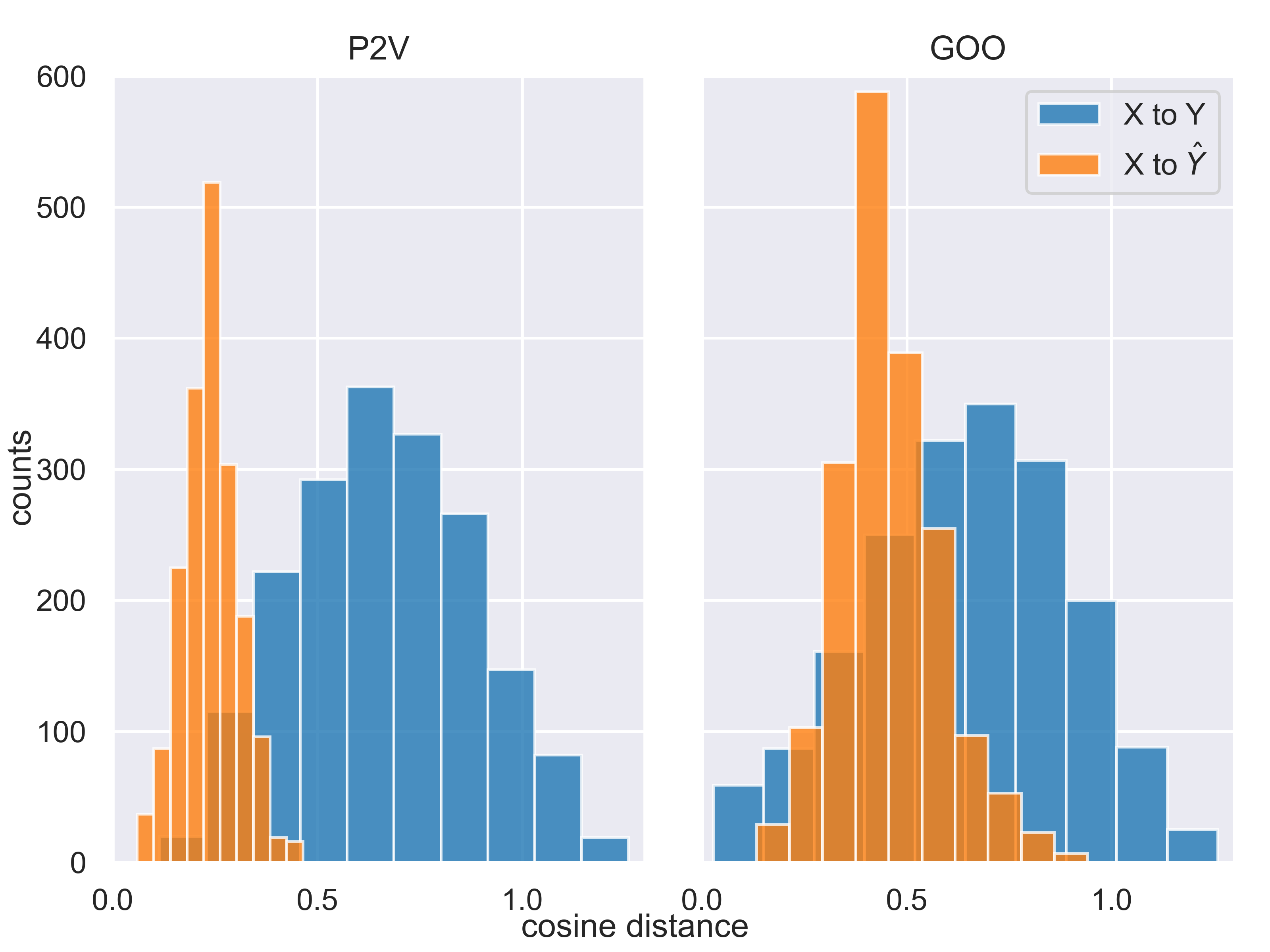}
  \vspace{-4mm}
  \caption{Distribution of cosine distances for input to label (X to Y, blue) and input to prediction (X to $\hat{Y}$, orange).}
  \label{fig:query_dist}
 \vspace{-5mm}
\end{figure}

To showcase \texttt{RecList} in a real-world setting, we test three RSs on a complementary items task: a \textit{prod2vec}-based recommender \cite{Bianchi2020FantasticEA} (hence \textbf{P2V}); Google Recommendation APIs (\textbf{GOO}) \cite{goorecs}; and one popular SaaS model (\textbf{S1})\footnote{Due to monetary and legal limitations, a perfect comparison on our private dataset was impossible. The goal of this analysis is \textit{not} to rank these models, but to demonstrate how the methodology provides insights about their behavior.}. We use data from a ``reasonable scale'' \cite{10.1145/3460231.3474604} e-commerce in the sport apparel industry, where 1M product interactions have been sampled for training from a period of 3 months in 2019, and 2.5K samples from a disjoint time period for testing. The main take-away of this experiment is simple: models (\textbf{GOO} and \textbf{P2V}) that are close when point-wise metrics are reported (Table~\ref{tab:results}) may have a very different behavior, when analyzed through \texttt{RecList}. In particular, we discuss three insightful \textit{RecTests} we performed:

\begin{itemize}
    \item \textbf{Product Popularity}: we compare model hits across item popularity (i.e. how accurate the prediction is, when the target is very / mildly / poorly popular). \textbf{P2V} can be seen to perform better on rare items by $40\%$ over \textbf{GOO}. On the other hand \textbf{GOO} outperforms \textbf{P2V} by $200\%$ on the most frequently-viewed items.  

    \item \textbf{``Being Less Wrong''}: we compute the cosine-distance (over a \textit{prod2vec} space) between query and ground truth, and query and prediction \textit{for missed predictions} (Figure~\ref{fig:query_dist}). We observe that  \textbf{GOO}'s prediction distribution better matches the label distribution, suggesting that its predictions are qualitatively more aligned to the complementary nature of the cart recommendation task \footnote{Qualitative checks confirmed that P2V often predicted products from the same category as input, whereas GOO exhibited greater prediction variety.}.
    
    \item \textbf{Slice-by-Brand}: we measure hits across various brands. While \textbf{P2V} and \textbf{GOO} have very similar overall performance, \textbf{P2V} is particularly performant on \texttt{asics}, compensating for a slightly lower result on \texttt{nike}: without behavioral testing, this bias in \textbf{P2V} would have been hard or time-consuming to catch.
\end{itemize}

Additional \textit{RecTests} are included in Table \ref{tab:results}: in particular, ``Being Less Wrong'' can be operationalized over brand affinity as well (\textit{Cos Distance (Brand)}), capturing the intuition that an Adidas product is closer to a Nike one than a Lacoste one. Conversely, \textit{Path Length} goes for a discrete approach and measures distance as the path length between input and prediction based on a product tree (longer suggests greater diversity, better for cart recommendations).

\section{Conclusion}
We introduced \texttt{RecList}, a package for behavioral testing in recommender systems. \texttt{RecList} aims to both provide a shared lexicon to explicitly discuss RSs trade-offs, and a convenient API for scaling and re-use behavioral tests. Our \textit{alpha} already provides out of the box support for popular datasets and common tests; not dissimilarly from Lego blocks, existing lists can be extended with new tests, tests can be re-assembled for different purposes, and -- as long as ``blocks'' implement the proper interface -- entirely new \textit{RecLists} can be created. We \textit{are} indeed aware that \texttt{RecList} is, by nature, a never-ending and continuously improving project: behavioral testing needs to constantly evolve as our understanding of RSs improves and their capabilities and reach change: by open sourcing \texttt{RecList}, we hope to help the field go beyond ``leaderboard chasing'', and to empower practitioners with better tools for analysis, debugging, and decision-making. 

\begin{acks}
Authors wish to thank three anonymous reviewers, Andrea Polonioli and Ciro Greco for feedback on previous drafts of this work and Jean-Francis Roy for his constant support in this project (and many others as well). Finally, it is worth mentioning that this is our first community-sourced (5 cities, 4 time zones) scholarly work, with Chloe and Brian joining the \textit{Coveo} line-up through a thread on Discord (thanks Chip Huyen for creating that amazing place!). While it is too early to say how successful this model will be for a company of our size, we are proud of what we achieved so far.
\end{acks}

\bibliographystyle{ACM-Reference-Format}
\bibliography{sample-base}


\begin{thebibliography}{40}

%%% the Web address.

\ifx \showCODEN    \undefined \def \showCODEN     #1{\unskip}     \fi
\ifx \showDOI      \undefined \def \showDOI       #1{#1}\fi
\ifx \showISBNx    \undefined \def \showISBNx     #1{\unskip}     \fi
\ifx \showISBNxiii \undefined \def \showISBNxiii  #1{\unskip}     \fi
\ifx \showISSN     \undefined \def \showISSN      #1{\unskip}     \fi
\ifx \showLCCN     \undefined \def \showLCCN      #1{\unskip}     \fi
\ifx \shownote     \undefined \def \shownote      #1{#1}          \fi
\ifx \showarticletitle \undefined \def \showarticletitle #1{#1}   \fi
\ifx \showURL      \undefined \def \showURL       {\relax}        \fi
% invisible to TeX
\providecommand\bibfield[2]{#2}
\providecommand\bibinfo[2]{#2}
\providecommand\natexlab[1]{#1}
\providecommand\showeprint[2][]{arXiv:#2}

\bibitem[\protect\citeauthoryear{Adamopoulos and Tuzhilin}{Adamopoulos and
  Tuzhilin}{2014}]%
        {concentrationbias}
\bibfield{author}{\bibinfo{person}{Panagiotis Adamopoulos} {and}
  \bibinfo{person}{Alexander Tuzhilin}.} \bibinfo{year}{2014}\natexlab{}.
\newblock \showarticletitle{On Over-Specialization and Concentration Bias of
  Recommendations: Probabilistic Neighborhood Selection in Collaborative
  Filtering Systems}. In \bibinfo{booktitle}{\emph{Proceedings of the 8th ACM
  Conference on Recommender Systems}} (Foster City, Silicon Valley, California,
  USA) \emph{(\bibinfo{series}{RecSys '14})}. \bibinfo{publisher}{Association
  for Computing Machinery}, \bibinfo{address}{New York, NY, USA},
  \bibinfo{pages}{153–160}.
\newblock
\showISBNx{9781450326681}
\urldef\tempurl%
\url{https://doi.org/10.1145/2645710.2645752}
\showDOI{\tempurl}


\bibitem[\protect\citeauthoryear{Arora, Ensslen, Fiedler, Liu, Robinson, Stein,
  and Schüler}{Arora et~al\mbox{.}}{2021}]%
        {mckinsey}
\bibfield{author}{\bibinfo{person}{Nidhi Arora}, \bibinfo{person}{Daniel
  Ensslen}, \bibinfo{person}{Lars Fiedler}, \bibinfo{person}{Wei~Wei Liu},
  \bibinfo{person}{Kelsey Robinson}, \bibinfo{person}{Eli Stein}, {and}
  \bibinfo{person}{Gustavo Schüler}.} \bibinfo{year}{2021}\natexlab{}.
\newblock \bibinfo{booktitle}{\emph{The value of getting personalization
  right—or wrong—is multiplying}}.
\newblock
\urldef\tempurl%
\url{https://www.mckinsey.com/business-functions/marketing-and-sales/our-insights/the-value-of-getting-personalization-right-or-wrong-is-multiplying}
\showURL{%
Retrieved November 15, 2021 from \tempurl}


\bibitem[\protect\citeauthoryear{Beizer and Wiley}{Beizer and Wiley}{1996}]%
        {536464}
\bibfield{author}{\bibinfo{person}{B. Beizer} {and} \bibinfo{person}{J.
  Wiley}.} \bibinfo{year}{1996}\natexlab{}.
\newblock \showarticletitle{Black Box Testing: Techniques for Functional
  Testing of Software and Systems}.
\newblock \bibinfo{journal}{\emph{IEEE Software}} \bibinfo{volume}{13},
  \bibinfo{number}{5} (\bibinfo{year}{1996}), \bibinfo{pages}{98--}.
\newblock
\urldef\tempurl%
\url{https://doi.org/10.1109/MS.1996.536464}
\showDOI{\tempurl}


\bibitem[\protect\citeauthoryear{Berg, Chirravuri, Cledat, Goyal, Hamad, and
  Tuulos}{Berg et~al\mbox{.}}{2019}]%
        {Metaflow}
\bibfield{author}{\bibinfo{person}{David Berg}, \bibinfo{person}{Ravi~Kiran
  Chirravuri}, \bibinfo{person}{Romain Cledat}, \bibinfo{person}{Savin Goyal},
  \bibinfo{person}{Ferras Hamad}, {and} \bibinfo{person}{Ville Tuulos}.}
  \bibinfo{year}{2019}\natexlab{}.
\newblock \bibinfo{booktitle}{\emph{Open-Sourcing Metaflow, a Human-Centric
  Framework for Data Science}}.
\newblock
\urldef\tempurl%
\url{https://netflixtechblog.com/open-sourcing-metaflow-a-human-centric-framework-for-data-science-fa72e04a5d9}
\showURL{%
\tempurl}


\bibitem[\protect\citeauthoryear{Bhagat, Muralidharan, Lobzhanidze, and
  Vishwanath}{Bhagat et~al\mbox{.}}{2018}]%
        {10.1145/3219819.3219891}
\bibfield{author}{\bibinfo{person}{Rahul Bhagat}, \bibinfo{person}{Srevatsan
  Muralidharan}, \bibinfo{person}{Alex Lobzhanidze}, {and}
  \bibinfo{person}{Shankar Vishwanath}.} \bibinfo{year}{2018}\natexlab{}.
\newblock \showarticletitle{Buy It Again: Modeling Repeat Purchase
  Recommendations}. In \bibinfo{booktitle}{\emph{Proceedings of the 24th ACM
  SIGKDD International Conference on Knowledge Discovery and Data Mining}}
  (London, United Kingdom) \emph{(\bibinfo{series}{KDD '18})}.
  \bibinfo{publisher}{Association for Computing Machinery},
  \bibinfo{address}{New York, NY, USA}, \bibinfo{pages}{62–70}.
\newblock
\showISBNx{9781450355520}
\urldef\tempurl%
\url{https://doi.org/10.1145/3219819.3219891}
\showDOI{\tempurl}


\bibitem[\protect\citeauthoryear{Bianchi, Tagliabue, Yu, Bigon, and
  Greco}{Bianchi et~al\mbox{.}}{2020}]%
        {Bianchi2020FantasticEA}
\bibfield{author}{\bibinfo{person}{Federico Bianchi}, \bibinfo{person}{J.
  Tagliabue}, \bibinfo{person}{Bingqing Yu}, \bibinfo{person}{Luca Bigon},
  {and} \bibinfo{person}{Ciro Greco}.} \bibinfo{year}{2020}\natexlab{}.
\newblock \showarticletitle{Fantastic Embeddings and How to Align Them:
  Zero-Shot Inference in a Multi-Shop Scenario}.
\newblock \bibinfo{journal}{\emph{ArXiv}}  \bibinfo{volume}{abs/2007.14906}
  (\bibinfo{year}{2020}).
\newblock


\bibitem[\protect\citeauthoryear{Bianchi, Yu, and Tagliabue}{Bianchi
  et~al\mbox{.}}{2021}]%
        {tagliabue-etal-2021-bert}
\bibfield{author}{\bibinfo{person}{Federico Bianchi}, \bibinfo{person}{Bingqing
  Yu}, {and} \bibinfo{person}{Jacopo Tagliabue}.}
  \bibinfo{year}{2021}\natexlab{}.
\newblock \showarticletitle{{BERT} Goes Shopping: Comparing Distributional
  Models for Product Representations}. In \bibinfo{booktitle}{\emph{Proceedings
  of The 4th Workshop on e-Commerce and NLP}}. \bibinfo{publisher}{Association
  for Computational Linguistics}, \bibinfo{address}{Online},
  \bibinfo{pages}{1--12}.
\newblock
\urldef\tempurl%
\url{https://doi.org/10.18653/v1/2021.ecnlp-1.1}
\showDOI{\tempurl}


\bibitem[\protect\citeauthoryear{Cai, Wu, San, Wang, and Wang}{Cai
  et~al\mbox{.}}{2021}]%
        {10.1145/3404835.3462832}
\bibfield{author}{\bibinfo{person}{Renqin Cai}, \bibinfo{person}{Jibang Wu},
  \bibinfo{person}{Aidan San}, \bibinfo{person}{Chong Wang}, {and}
  \bibinfo{person}{Hongning Wang}.} \bibinfo{year}{2021}\natexlab{}.
\newblock \showarticletitle{Category-Aware Collaborative Sequential
  Recommendation}. In \bibinfo{booktitle}{\emph{Proceedings of the 44th
  International ACM SIGIR Conference on Research and Development in Information
  Retrieval}} (Virtual Event, Canada) \emph{(\bibinfo{series}{SIGIR '21})}.
  \bibinfo{publisher}{Association for Computing Machinery},
  \bibinfo{address}{New York, NY, USA}, \bibinfo{pages}{388–397}.
\newblock
\showISBNx{9781450380379}
\urldef\tempurl%
\url{https://doi.org/10.1145/3404835.3462832}
\showDOI{\tempurl}


\bibitem[\protect\citeauthoryear{Chia, Yu, and Tagliabue}{Chia
  et~al\mbox{.}}{2021}]%
        {Chia2021AreYS}
\bibfield{author}{\bibinfo{person}{Patrick~John Chia},
  \bibinfo{person}{Bingqing Yu}, {and} \bibinfo{person}{Jacopo Tagliabue}.}
  \bibinfo{year}{2021}\natexlab{}.
\newblock \showarticletitle{"Are you sure?": Preliminary Insights from Scaling
  Product Comparisons to Multiple Shops}.
\newblock \bibinfo{journal}{\emph{ArXiv}}  \bibinfo{volume}{abs/2107.03256}
  (\bibinfo{year}{2021}).
\newblock


\bibitem[\protect\citeauthoryear{Cloud}{Cloud}{2021}]%
        {goorecs}
\bibfield{author}{\bibinfo{person}{Google Cloud}.}
  \bibinfo{year}{2021}\natexlab{}.
\newblock \bibinfo{booktitle}{\emph{Implementing Recommendations AI}}.
\newblock
\urldef\tempurl%
\url{https://cloud.google.com/retail/recommendations-ai/docs/overview}
\showURL{%
Retrieved November 17, 2021 from \tempurl}


\bibitem[\protect\citeauthoryear{Comet.ML}{Comet.ML}{2021}]%
        {CometML}
\bibfield{author}{\bibinfo{person}{Comet.ML}.} \bibinfo{year}{2021}\natexlab{}.
\newblock \bibinfo{booktitle}{\emph{{Comet.ML} home page}}.
\newblock
\urldef\tempurl%
\url{https://www.comet.ml/}
\showURL{%
\tempurl}


\bibitem[\protect\citeauthoryear{de~Souza Pereira~Moreira, Rabhi, Ak, Kabir,
  and Oldridge}{de~Souza Pereira~Moreira et~al\mbox{.}}{2021}]%
        {Moreira2021TransformersWM}
\bibfield{author}{\bibinfo{person}{Gabriel de Souza Pereira~Moreira},
  \bibinfo{person}{Sara Rabhi}, \bibinfo{person}{Ronay Ak},
  \bibinfo{person}{Md~Yasin Kabir}, {and} \bibinfo{person}{Even Oldridge}.}
  \bibinfo{year}{2021}\natexlab{}.
\newblock \showarticletitle{Transformers with multi-modal features and
  post-fusion context for e-commerce session-based recommendation}.
\newblock \bibinfo{journal}{\emph{ArXiv}}  \bibinfo{volume}{abs/2107.05124}
  (\bibinfo{year}{2021}).
\newblock


\bibitem[\protect\citeauthoryear{Guo, Yin, Wang, Chen, Zhou, and Quoc
  Viet~Hung}{Guo et~al\mbox{.}}{2019}]%
        {10.1145/3292500.3330839}
\bibfield{author}{\bibinfo{person}{Lei Guo}, \bibinfo{person}{Hongzhi Yin},
  \bibinfo{person}{Qinyong Wang}, \bibinfo{person}{Tong Chen},
  \bibinfo{person}{Alexander Zhou}, {and} \bibinfo{person}{Nguyen Quoc
  Viet~Hung}.} \bibinfo{year}{2019}\natexlab{}.
\newblock \showarticletitle{Streaming Session-Based Recommendation}. In
  \bibinfo{booktitle}{\emph{Proceedings of the 25th ACM SIGKDD International
  Conference on Knowledge Discovery and Data Mining}} (Anchorage, AK, USA)
  \emph{(\bibinfo{series}{KDD '19})}. \bibinfo{publisher}{Association for
  Computing Machinery}, \bibinfo{address}{New York, NY, USA},
  \bibinfo{pages}{1569–1577}.
\newblock
\showISBNx{9781450362016}
\urldef\tempurl%
\url{https://doi.org/10.1145/3292500.3330839}
\showDOI{\tempurl}


\bibitem[\protect\citeauthoryear{Harper and Konstan}{Harper and
  Konstan}{2015}]%
        {10.1145/2827872}
\bibfield{author}{\bibinfo{person}{F.~Maxwell Harper} {and}
  \bibinfo{person}{Joseph~A. Konstan}.} \bibinfo{year}{2015}\natexlab{}.
\newblock \showarticletitle{The MovieLens Datasets: History and Context}.
\newblock \bibinfo{journal}{\emph{ACM Trans. Interact. Intell. Syst.}}
  \bibinfo{volume}{5}, \bibinfo{number}{4}, Article \bibinfo{articleno}{19}
  (\bibinfo{date}{Dec.} \bibinfo{year}{2015}), \bibinfo{numpages}{19}~pages.
\newblock
\showISSN{2160-6455}
\urldef\tempurl%
\url{https://doi.org/10.1145/2827872}
\showDOI{\tempurl}


\bibitem[\protect\citeauthoryear{Hidasi, Karatzoglou, Baltrunas, and
  Tikk}{Hidasi et~al\mbox{.}}{2016}]%
        {Hidasi2016SessionbasedRW}
\bibfield{author}{\bibinfo{person}{Bal{\'a}zs Hidasi},
  \bibinfo{person}{Alexandros Karatzoglou}, \bibinfo{person}{Linas Baltrunas},
  {and} \bibinfo{person}{Domonkos Tikk}.} \bibinfo{year}{2016}\natexlab{}.
\newblock \showarticletitle{Session-based Recommendations with Recurrent Neural
  Networks}.
\newblock \bibinfo{journal}{\emph{CoRR}}  \bibinfo{volume}{abs/1511.06939}
  (\bibinfo{year}{2016}).
\newblock


\bibitem[\protect\citeauthoryear{Hosanagar, Fleder, Lee, and Buja}{Hosanagar
  et~al\mbox{.}}{2014}]%
        {netflixarticle}
\bibfield{author}{\bibinfo{person}{Kartik Hosanagar}, \bibinfo{person}{Daniel
  Fleder}, \bibinfo{person}{Dokyun Lee}, {and} \bibinfo{person}{Andreas Buja}.}
  \bibinfo{year}{2014}\natexlab{}.
\newblock \showarticletitle{Will the Global Village Fracture Into Tribes?
  Recommender Systems and Their Effects on Consumer Fragmentation}.
\newblock \bibinfo{journal}{\emph{Management Science}}  \bibinfo{volume}{60}
  (\bibinfo{date}{04} \bibinfo{year}{2014}), \bibinfo{pages}{805--823}.
\newblock
\urldef\tempurl%
\url{https://doi.org/10.1287/mnsc.2013.1808}
\showDOI{\tempurl}


\bibitem[\protect\citeauthoryear{Jannach and Ludewig}{Jannach and
  Ludewig}{2017}]%
        {jannach2017recurrent}
\bibfield{author}{\bibinfo{person}{Dietmar Jannach} {and}
  \bibinfo{person}{Malte Ludewig}.} \bibinfo{year}{2017}\natexlab{}.
\newblock \showarticletitle{When recurrent neural networks meet the
  neighborhood for session-based recommendation}. In
  \bibinfo{booktitle}{\emph{Proceedings of the Eleventh ACM Conference on
  Recommender Systems}}. \bibinfo{pages}{306--310}.
\newblock


\bibitem[\protect\citeauthoryear{Kohavi, Deng, Frasca, Longbotham, Walker, and
  Xu}{Kohavi et~al\mbox{.}}{2012}]%
        {10.1145/2339530.2339653}
\bibfield{author}{\bibinfo{person}{Ron Kohavi}, \bibinfo{person}{Alex Deng},
  \bibinfo{person}{Brian Frasca}, \bibinfo{person}{Roger Longbotham},
  \bibinfo{person}{Toby Walker}, {and} \bibinfo{person}{Ya Xu}.}
  \bibinfo{year}{2012}\natexlab{}.
\newblock \showarticletitle{Trustworthy Online Controlled Experiments: Five
  Puzzling Outcomes Explained}. In \bibinfo{booktitle}{\emph{Proceedings of the
  18th ACM SIGKDD International Conference on Knowledge Discovery and Data
  Mining}} (Beijing, China) \emph{(\bibinfo{series}{KDD '12})}.
  \bibinfo{publisher}{Association for Computing Machinery},
  \bibinfo{address}{New York, NY, USA}, \bibinfo{pages}{786–794}.
\newblock
\showISBNx{9781450314626}
\urldef\tempurl%
\url{https://doi.org/10.1145/2339530.2339653}
\showDOI{\tempurl}


\bibitem[\protect\citeauthoryear{Kotkov, Veijalainen, and Wang}{Kotkov
  et~al\mbox{.}}{2016}]%
        {Kotkov2016ChallengesOS}
\bibfield{author}{\bibinfo{person}{Denis Kotkov}, \bibinfo{person}{Jari
  Veijalainen}, {and} \bibinfo{person}{Shuaiqiang Wang}.}
  \bibinfo{year}{2016}\natexlab{}.
\newblock \showarticletitle{Challenges of Serendipity in Recommender Systems}.
  In \bibinfo{booktitle}{\emph{WEBIST}}.
\newblock


\bibitem[\protect\citeauthoryear{Kouki, Fountalis, Vasiloglou, Cui, Liberty,
  and Al~Jadda}{Kouki et~al\mbox{.}}{2020}]%
        {10.1145/3383313.3412235}
\bibfield{author}{\bibinfo{person}{Pigi Kouki}, \bibinfo{person}{Ilias
  Fountalis}, \bibinfo{person}{Nikolaos Vasiloglou}, \bibinfo{person}{Xiquan
  Cui}, \bibinfo{person}{Edo Liberty}, {and} \bibinfo{person}{Khalifeh
  Al~Jadda}.} \bibinfo{year}{2020}\natexlab{}.
\newblock \showarticletitle{From the Lab to Production: A Case Study of
  Session-Based Recommendations in the Home-Improvement Domain}. In
  \bibinfo{booktitle}{\emph{Fourteenth ACM Conference on Recommender Systems}}
  (Virtual Event, Brazil) \emph{(\bibinfo{series}{RecSys '20})}.
  \bibinfo{publisher}{Association for Computing Machinery},
  \bibinfo{address}{New York, NY, USA}, \bibinfo{pages}{140–149}.
\newblock
\showISBNx{9781450375832}
\urldef\tempurl%
\url{https://doi.org/10.1145/3383313.3412235}
\showDOI{\tempurl}


\bibitem[\protect\citeauthoryear{{Krista Garcia}}{{Krista Garcia}}{2018}]%
        {emarketer}
\bibfield{author}{\bibinfo{person}{{Krista Garcia}}.}
  \bibinfo{year}{2018}\natexlab{}.
\newblock \bibinfo{booktitle}{\emph{The Impact of Product Recommendations}}.
\newblock
\urldef\tempurl%
\url{https://www.emarketer.com/content/the-impact-of-product-recommendations}
\showURL{%
Retrieved November 9, 2021 from \tempurl}


\bibitem[\protect\citeauthoryear{Lamkhede and Kofler}{Lamkhede and
  Kofler}{2021}]%
        {lamkhede2021recommendations}
\bibfield{author}{\bibinfo{person}{Sudarshan Lamkhede} {and}
  \bibinfo{person}{Christoph Kofler}.} \bibinfo{year}{2021}\natexlab{}.
\newblock \showarticletitle{Recommendations and Results Organization in Netflix
  Search}. In \bibinfo{booktitle}{\emph{RecSys '21: Fifteenth ACM Conference on
  Recommender Systems}}.
\newblock


\bibitem[\protect\citeauthoryear{Ludewig and Jannach}{Ludewig and
  Jannach}{2018}]%
        {ludewig2018evaluation}
\bibfield{author}{\bibinfo{person}{Malte Ludewig} {and}
  \bibinfo{person}{Dietmar Jannach}.} \bibinfo{year}{2018}\natexlab{}.
\newblock \showarticletitle{Evaluation of session-based recommendation
  algorithms}.
\newblock \bibinfo{journal}{\emph{User Modeling and User-Adapted Interaction}}
  \bibinfo{volume}{28}, \bibinfo{number}{4-5} (\bibinfo{year}{2018}),
  \bibinfo{pages}{331--390}.
\newblock


\bibitem[\protect\citeauthoryear{Mikolov, Chen, Corrado, and Dean}{Mikolov
  et~al\mbox{.}}{2013}]%
        {Mikolov2013EfficientEO}
\bibfield{author}{\bibinfo{person}{Tomas Mikolov}, \bibinfo{person}{Kai Chen},
  \bibinfo{person}{Gregory~S. Corrado}, {and} \bibinfo{person}{Jeffrey Dean}.}
  \bibinfo{year}{2013}\natexlab{}.
\newblock \showarticletitle{Efficient Estimation of Word Representations in
  Vector Space}. In \bibinfo{booktitle}{\emph{ICLR}}.
\newblock


\bibitem[\protect\citeauthoryear{Moins, Aloise, and Blanchard}{Moins
  et~al\mbox{.}}{2020}]%
        {10.1145/3383313.3412263}
\bibfield{author}{\bibinfo{person}{Th\'{e}o Moins}, \bibinfo{person}{Daniel
  Aloise}, {and} \bibinfo{person}{Simon~J. Blanchard}.}
  \bibinfo{year}{2020}\natexlab{}.
\newblock \showarticletitle{RecSeats: A Hybrid Convolutional Neural Network
  Choice Model for Seat Recommendations at Reserved Seating Venues}. In
  \bibinfo{booktitle}{\emph{Fourteenth ACM Conference on Recommender Systems}}
  (Virtual Event, Brazil) \emph{(\bibinfo{series}{RecSys '20})}.
  \bibinfo{publisher}{Association for Computing Machinery},
  \bibinfo{address}{New York, NY, USA}, \bibinfo{pages}{309–317}.
\newblock
\showISBNx{9781450375832}
\urldef\tempurl%
\url{https://doi.org/10.1145/3383313.3412263}
\showDOI{\tempurl}


\bibitem[\protect\citeauthoryear{Nassif, Cansizlar, Goodman, and
  Vishwanathan}{Nassif et~al\mbox{.}}{2018}]%
        {nassif2018diversifying}
\bibfield{author}{\bibinfo{person}{Houssam Nassif}, \bibinfo{person}{Kemal~Oral
  Cansizlar}, \bibinfo{person}{Mitchell Goodman}, {and} \bibinfo{person}{SVN
  Vishwanathan}.} \bibinfo{year}{2018}\natexlab{}.
\newblock \showarticletitle{Diversifying Music Recommendations}. In
  \bibinfo{booktitle}{\emph{ICML '16 Workshop}}.
\newblock


\bibitem[\protect\citeauthoryear{Radford, Kim, Hallacy, Ramesh, Goh, Agarwal,
  Sastry, Askell, Mishkin, Clark, Krueger, and Sutskever}{Radford
  et~al\mbox{.}}{2021}]%
        {Radford2021LearningTV}
\bibfield{author}{\bibinfo{person}{Alec Radford}, \bibinfo{person}{Jong~Wook
  Kim}, \bibinfo{person}{Chris Hallacy}, \bibinfo{person}{Aditya Ramesh},
  \bibinfo{person}{Gabriel Goh}, \bibinfo{person}{Sandhini Agarwal},
  \bibinfo{person}{Girish Sastry}, \bibinfo{person}{Amanda Askell},
  \bibinfo{person}{Pamela Mishkin}, \bibinfo{person}{Jack Clark},
  \bibinfo{person}{Gretchen Krueger}, {and} \bibinfo{person}{Ilya Sutskever}.}
  \bibinfo{year}{2021}\natexlab{}.
\newblock \showarticletitle{Learning Transferable Visual Models From Natural
  Language Supervision}. In \bibinfo{booktitle}{\emph{ICML}}.
\newblock


\bibitem[\protect\citeauthoryear{Rashed, Jawed, Schmidt-Thieme, and
  Hintsches}{Rashed et~al\mbox{.}}{2020}]%
        {Rashed2020MultiRecAM}
\bibfield{author}{\bibinfo{person}{Ahmed Rashed}, \bibinfo{person}{Shayan
  Jawed}, \bibinfo{person}{Lars Schmidt-Thieme}, {and} \bibinfo{person}{Andre
  Hintsches}.} \bibinfo{year}{2020}\natexlab{}.
\newblock \showarticletitle{MultiRec: A Multi-Relational Approach for Unique
  Item Recommendation in Auction Systems}.
\newblock \bibinfo{journal}{\emph{Fourteenth ACM Conference on Recommender
  Systems}} (\bibinfo{year}{2020}).
\newblock


\bibitem[\protect\citeauthoryear{Ribeiro, Wu, Guestrin, and Singh}{Ribeiro
  et~al\mbox{.}}{2020}]%
        {Ribeiro2020BeyondAB}
\bibfield{author}{\bibinfo{person}{Marco~T{\'u}lio Ribeiro},
  \bibinfo{person}{Tongshuang~(Sherry) Wu}, \bibinfo{person}{Carlos Guestrin},
  {and} \bibinfo{person}{Sameer Singh}.} \bibinfo{year}{2020}\natexlab{}.
\newblock \showarticletitle{Beyond Accuracy: Behavioral Testing of NLP Models
  with CheckList}. In \bibinfo{booktitle}{\emph{ACL}}.
\newblock


\bibitem[\protect\citeauthoryear{Saberian and Basilico}{Saberian and
  Basilico}{2021}]%
        {10.1145/3460231.3474620}
\bibfield{author}{\bibinfo{person}{Mohammad Saberian} {and}
  \bibinfo{person}{Justin Basilico}.} \bibinfo{year}{2021}\natexlab{}.
\newblock \bibinfo{booktitle}{\emph{RecSysOps: Best Practices for Operating a
  Large-Scale Recommender System}}.
\newblock \bibinfo{publisher}{Association for Computing Machinery},
  \bibinfo{address}{New York, NY, USA}, \bibinfo{pages}{590–591}.
\newblock
\showISBNx{9781450384582}
\urldef\tempurl%
\url{https://doi.org/10.1145/3460231.3474620}
\showURL{%
\tempurl}


\bibitem[\protect\citeauthoryear{{Statista Research Department}}{{Statista
  Research Department}}{2020}]%
        {ecomworld}
\bibfield{author}{\bibinfo{person}{{Statista Research Department}}.}
  \bibinfo{year}{2020}\natexlab{}.
\newblock \bibinfo{booktitle}{\emph{Global retail e-commerce sales 2014-2023}}.
\newblock
\urldef\tempurl%
\url{https://www.statista.com/statistics/379046/worldwide-retail-e-commerce-sales/}
\showURL{%
Retrieved November 29, 2020 from \tempurl}


\bibitem[\protect\citeauthoryear{Tagliabue}{Tagliabue}{2021}]%
        {10.1145/3460231.3474604}
\bibfield{author}{\bibinfo{person}{Jacopo Tagliabue}.}
  \bibinfo{year}{2021}\natexlab{}.
\newblock \bibinfo{booktitle}{\emph{You Do Not Need a Bigger Boat:
  Recommendations at Reasonable Scale in a (Mostly) Serverless and Open
  Stack}}.
\newblock \bibinfo{publisher}{Association for Computing Machinery},
  \bibinfo{address}{New York, NY, USA}, \bibinfo{pages}{598–600}.
\newblock
\showISBNx{9781450384582}
\urldef\tempurl%
\url{https://doi.org/10.1145/3460231.3474604}
\showURL{%
\tempurl}


\bibitem[\protect\citeauthoryear{Tagliabue, Greco, Roy, Bianchi, Cassani, Yu,
  and Chia}{Tagliabue et~al\mbox{.}}{2021}]%
        {CoveoSIGIR2021}
\bibfield{author}{\bibinfo{person}{Jacopo Tagliabue}, \bibinfo{person}{Ciro
  Greco}, \bibinfo{person}{Jean-Francis Roy}, \bibinfo{person}{Federico
  Bianchi}, \bibinfo{person}{Giovanni Cassani}, \bibinfo{person}{Bingqing Yu},
  {and} \bibinfo{person}{Patrick~John Chia}.} \bibinfo{year}{2021}\natexlab{}.
\newblock \showarticletitle{SIGIR 2021 E-Commerce Workshop Data Challenge}. In
  \bibinfo{booktitle}{\emph{SIGIR eCom 2021}}.
\newblock


\bibitem[\protect\citeauthoryear{Tagliabue, Yu, and Bianchi}{Tagliabue
  et~al\mbox{.}}{2020}]%
        {10.1145/3383313.3411477}
\bibfield{author}{\bibinfo{person}{Jacopo Tagliabue}, \bibinfo{person}{Bingqing
  Yu}, {and} \bibinfo{person}{Federico Bianchi}.}
  \bibinfo{year}{2020}\natexlab{}.
\newblock \showarticletitle{The Embeddings That Came in From the Cold:
  Improving Vectors for New and Rare Products with Content-Based Inference}. In
  \bibinfo{booktitle}{\emph{Fourteenth ACM Conference on Recommender Systems}}
  (Virtual Event, Brazil) \emph{(\bibinfo{series}{RecSys '20})}.
  \bibinfo{publisher}{Association for Computing Machinery},
  \bibinfo{address}{New York, NY, USA}, \bibinfo{pages}{577–578}.
\newblock
\showISBNx{9781450375832}
\urldef\tempurl%
\url{https://doi.org/10.1145/3383313.3411477}
\showDOI{\tempurl}


\bibitem[\protect\citeauthoryear{Vasile, Smirnova, and Conneau}{Vasile
  et~al\mbox{.}}{2016}]%
        {10.1145/2959100.2959160}
\bibfield{author}{\bibinfo{person}{Flavian Vasile}, \bibinfo{person}{Elena
  Smirnova}, {and} \bibinfo{person}{Alexis Conneau}.}
  \bibinfo{year}{2016}\natexlab{}.
\newblock \showarticletitle{Meta-Prod2Vec: Product Embeddings Using
  Side-Information for Recommendation}. In
  \bibinfo{booktitle}{\emph{Proceedings of the 10th ACM Conference on
  Recommender Systems}} (Boston, Massachusetts, USA)
  \emph{(\bibinfo{series}{RecSys '16})}. \bibinfo{publisher}{Association for
  Computing Machinery}, \bibinfo{address}{New York, NY, USA},
  \bibinfo{pages}{225–232}.
\newblock
\showISBNx{9781450340359}
\urldef\tempurl%
\url{https://doi.org/10.1145/2959100.2959160}
\showDOI{\tempurl}


\bibitem[\protect\citeauthoryear{Wang, Lin, Lin, Yang, and Wu}{Wang
  et~al\mbox{.}}{2020b}]%
        {DBLP:journals/corr/abs-2005-10110}
\bibfield{author}{\bibinfo{person}{Menghan Wang}, \bibinfo{person}{Yujie Lin},
  \bibinfo{person}{Guli Lin}, \bibinfo{person}{Keping Yang}, {and}
  \bibinfo{person}{Xiao{-}Ming Wu}.} \bibinfo{year}{2020}\natexlab{b}.
\newblock \showarticletitle{{M2GRL:} {A} Multi-task Multi-view Graph
  Representation Learning Framework for Web-scale Recommender Systems}.
\newblock \bibinfo{journal}{\emph{CoRR}}  \bibinfo{volume}{abs/2005.10110}
  (\bibinfo{year}{2020}).
\newblock
\showeprint[arxiv]{2005.10110}
\urldef\tempurl%
\url{https://arxiv.org/abs/2005.10110}
\showURL{%
\tempurl}


\bibitem[\protect\citeauthoryear{Wang, Cao, and Wang}{Wang
  et~al\mbox{.}}{2019a}]%
        {Wang2019ASO}
\bibfield{author}{\bibinfo{person}{Shoujin Wang}, \bibinfo{person}{Longbing
  Cao}, {and} \bibinfo{person}{Yan Wang}.} \bibinfo{year}{2019}\natexlab{a}.
\newblock \showarticletitle{A Survey on Session-based Recommender Systems}.
\newblock \bibinfo{journal}{\emph{ACM Computing Surveys (CSUR)}}
  \bibinfo{volume}{54} (\bibinfo{year}{2019}), \bibinfo{pages}{1 -- 38}.
\newblock


\bibitem[\protect\citeauthoryear{Wang, He, Wang, Feng, and Chua}{Wang
  et~al\mbox{.}}{2019b}]%
        {wang2019neural}
\bibfield{author}{\bibinfo{person}{Xiang Wang}, \bibinfo{person}{Xiangnan He},
  \bibinfo{person}{Meng Wang}, \bibinfo{person}{Fuli Feng}, {and}
  \bibinfo{person}{Tat-Seng Chua}.} \bibinfo{year}{2019}\natexlab{b}.
\newblock \showarticletitle{Neural graph collaborative filtering}. In
  \bibinfo{booktitle}{\emph{Proceedings of the 42nd international ACM SIGIR
  conference on Research and development in Information Retrieval}}.
  \bibinfo{pages}{165--174}.
\newblock


\bibitem[\protect\citeauthoryear{Wang, Liang, Charlin, and Blei}{Wang
  et~al\mbox{.}}{2020a}]%
        {wang2019deconfounded}
\bibfield{author}{\bibinfo{person}{Yixin Wang}, \bibinfo{person}{Dawen Liang},
  \bibinfo{person}{Laurent Charlin}, {and} \bibinfo{person}{David~M. Blei}.}
  \bibinfo{year}{2020}\natexlab{a}.
\newblock \showarticletitle{Causal Inference for Recommender Systems}. In
  \bibinfo{booktitle}{\emph{recsys2020}}.
\newblock


\bibitem[\protect\citeauthoryear{Zamani, Schedl, Lamere, and Chen}{Zamani
  et~al\mbox{.}}{2019}]%
        {10.1145/3344257}
\bibfield{author}{\bibinfo{person}{Hamed Zamani}, \bibinfo{person}{Markus
  Schedl}, \bibinfo{person}{Paul Lamere}, {and} \bibinfo{person}{Ching-Wei
  Chen}.} \bibinfo{year}{2019}\natexlab{}.
\newblock \showarticletitle{An Analysis of Approaches Taken in the ACM RecSys
  Challenge 2018 for Automatic Music Playlist Continuation}.
\newblock \bibinfo{journal}{\emph{ACM Trans. Intell. Syst. Technol.}}
  \bibinfo{volume}{10}, \bibinfo{number}{5}, Article \bibinfo{articleno}{57}
  (\bibinfo{date}{Sept.} \bibinfo{year}{2019}), \bibinfo{numpages}{21}~pages.
\newblock
\showISSN{2157-6904}
\urldef\tempurl%
\url{https://doi.org/10.1145/3344257}
\showDOI{\tempurl}


\end{thebibliography}

%\appendix

%\section{Tracking and Visualization}
%\label{appendix:gui}

%Running a \textit{RecList}, given a \textit{RecDataset} and \textit{RecModel}, automatically produces a report, versioned by the list name, the model name and the timestamp of the run. Running different models (or the same one, trained with different parameters) will therefore produce separate reports, which can be listed and inspected with a small web application bundled with the package (Figure~\ref{fig:app}). 

%Selecting multiple reports results in out-of-the-box comparative tables and charts; being a static HTML, the page can be easily shared among all stakeholders. Since reporting and visualization are completely decoupled, it is possible for practitioners to either extend the visualization to encompass new capabilities, or to send the metrics as a machine-friendly payload to downstream systems.

%Please note that the \textit{alpha} version of \texttt{RecList} re-uses styling and code from the original Dag Cards \cite{Tagliabue2021DAGCI}.

%\begin{figure}[h]
%  \centering
%  \includegraphics[width=\linewidth]{app.jpg}
 % \caption{Built-in comparison and visualization app. Inspired by modern MLOps tools \cite{dbt}, \texttt{RecList} ships with a small web-app that prompts the user to select test runs (top), and then easily compare them through tables and distributions (bottom).}
%  \label{fig:app}
%\end{figure}

\end{document}